\documentstyle[aps,multicol,epsfig]{revtex}
\draft
\begin{document}

\title{Positive strangeness contribution to the nucleon magnetic moment 
	in a relativistic chiral potential model}
\author{X.B. Chen$^{a,b}$, X.S. Chen$^{a,c}$, Amand Faessler$^c$,
        Th. Gutsche$^c$, F. Wang$^a$}
\address{$^a$Department of Physics and Center for Theoretical Physics,
                Nanjing University, Nanjing 210093, China\\
        $^b$Department of Physics, Changsha Institute of Electricity,
                Changsha 410077, China\\
        $^c$Institut f\"ur Theoretische Physik, Universit\"at T\"ubingen,
                Auf der Morgenstelle 14, D-72076 T\"ubingen, Germany}
\date{March 15, 2000}
\maketitle

\begin{abstract}
The strangeness contribution to the nucleon magnetic moment is calculated 
at the one-loop level
in a relativistic SU(3) chiral potential model and is found to be 
{\em positive}, that is, with an {\em opposite} sign to the nucleon 
strangeness polarization.
It is the ``Z'' diagram that violates the usual relation between spin and 
magnetic moment. 
The positive value is due to the contribution from the intermediate excited 
quark states, while the intermediate  
ground state gives a negative contribution. Our numerical results agree 
quite well with the new measurement of the SAMPLE Collaboration.  

\end{abstract}

\begin{multicols}{2}
The new determination of the strangeness magnetic form factor of the nucleon  
at $Q^2=0.1$GeV$^2$ 
by the SAMPLE collaboration \cite{SAMPLE} confirmed its previous result 
\cite{SAMPLE0} and indicates a significantly {\em positive} value:
\begin{equation}
G^s_M(0.1{\rm GeV}^2)=+0.61\pm 0.17\pm 0.21 \pm 0.19\mu_N.
\end{equation}
This result implies new challenges to our understanding of the nucleon 
structure since 
most theoretical calculations typically generate negative values for 
$\mu_s\equiv G^s_M(Q^2=0)$ (see \cite{McKe,Dong} and references therein). 
A positive value for $\mu_s$ is also intuitively difficult to understand 
with respect to the usual magnetic moment-spin ($\mu$-$s$) relation, 
since the strange quark
polarization of the nucleon is confirmed to be {\em negative} both by
experiments \cite{SpinExp} and by lattice QCD calculations 
\cite{SpinLatt1,SpinLatt2}. (Note that the 
negative charge of the strange quark has been extracted in the definition of 
$G^s_M$.) 

In a recent paper \cite{Chen}, by using a relativistic chiral potential 
model, we have successfully reproduced the experimental result of
the strange quark polarization ($\Delta s$) of the nucleon. In this paper we
report on the corresponding result for the  nucleon strangeness 
magnetic moment, 
especially in comparison to the nucleon strangeness polarization. 

The nucleon magnetic moment $\mu_N $ is defined by its interaction with a
static, external magnetic field $\vec B$:
\begin{equation}
\langle N| \sum_q-iQ_q\int d^3x\bar \psi _q\gamma ^\mu
\psi _qA_\mu | N\rangle \equiv -\vec \mu_N \cdot \vec B,  \label{1}
\end{equation}
It can be shown that $\mu_N$ is related to the electromagnetic form factor 
by $\mu_N=\sum_q Q_qG_M^q(0)$ \cite{Wein}, 
where $G^q_M(k^2)\equiv F_1^q(k^2)+F_2^q(k^2)$ and
$F_1$ and $F_2$ are defined through
\begin{equation}
\langle N| \bar\psi _q \gamma^{\mu} \psi_q|N\rangle =
\bar u_N\left(F_1^q\gamma^{\mu}+
	      \frac{i}{2M_N}F_2^q\sigma^{\mu\nu}k_\nu\right)u_N.
\end{equation}
The contribution of the quark flavor $q$ ($q=u,d,s$) to the nucleon 
magnetic moment is usually defined as $\mu_q\equiv G^q_M(0)$. 
Equivalently, $\mu_q$ can be evaluated as the expectation value of 
the magnetic moment operator: 
\begin{equation}
\mu_q=\langle N| \int d^3x \psi^{\dagger}_q (\vec x \times \vec{\alpha})_3 
	\psi_q|N\rangle , \label{Def}
\end{equation}
which follows directly from Eq. (\ref{1}) \cite{Lee}. 
Eq. (\ref{Def}) is especially suitable for model calculations. In the following
we will perform a perturbative calculation of $\mu_s$ in a chiral potential 
model. 

Our starting point is the chiral Lagrangian
\begin{eqnarray}
{\cal L}&=&\bar{\psi} [ i\partial \hspace*{-2mm}/
                - S(r)- \gamma^0 V(r)]\psi - \nonumber \\
         &&\frac{1}{2F_\pi} \bar{\psi} [ 
        S(r)(\sigma +i\gamma ^5 \lambda^i \phi_i) + 
        (\sigma +i\gamma ^5\lambda^i \phi_i )S(r) ]\psi + \nonumber \\
         &&\frac 12 (\partial_\mu \sigma)^2 + 
           \frac 12 (\partial_\mu \phi_i)^2 -
           \frac 12 m_\sigma^2 \sigma^2 -
           \frac 12 m_{\phi_i}^2\phi_i^2.
            \label{Lagrangian} 
\end{eqnarray}
The model Lagrangian is derived from the $\sigma$ model in which meson fields
are introduced to restore chiral symmetry \cite{Thomas}. The flavor and color
indices for the quark field $\psi$ are suppressed; the scalar term
$S(r)=cr+m$ represents the the linear scalar confinement potential $cr$ and 
the quark
mass matrix $m$;
$V(r)=-\alpha /r$ is the Coulomb type vector potential and
$F_\pi$=93MeV is the pion decay
constant. $\sigma$ and $\phi_i$ ($i$ runs from $1$ to $8$) are the scalar
and pseudoscalar meson fields, respectively and $\lambda_i$ are the Gell-Mann
matrices. The quark-meson interaction term of Eq.(\ref{Lagrangian}) is
symmetrized since the 
mass matrix $m$ does not commute with all $\lambda_i$
for different quark masses.

At zeroth order the nucleon is described by the usual SU(6) three-quark ground 
state of the Hamiltonian
\begin{equation}
H_q = \int d^3 x \psi ^\dagger  [\vec \alpha \cdot \frac 1i\vec \partial + 
        \beta S(r) +V(r) ] \psi. 
\end{equation}
The lowest order contribution to $\mu_s$ arises from the one-loop 
diagrams of
Fig. 1, which we now evaluate. 

\begin{center}
\begin{minipage}{8.5cm}
\begin{figure}
\begin{center}
\psfig{figure=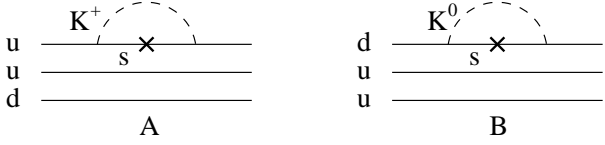,width=8cm}
\bigskip
\begin{tighten}
\caption{Lowest order diagram for $\mu_s$;
a cross on the quark line
 denotes the magnetic moment vertex  
$(\vec x \times \vec \alpha)_3$.}
\end{tighten}
\end{center}
\end{figure}
\end{minipage}
\end{center}

The meson propagator, given by the Lagrangian of Eq. (\ref{Lagrangian}), is 
the free propagator. 
Since the non-perturbative confinement is included in 
$H_q$ the quark propagator has to be obtained numerically, and in practise we 
have to work with time-ordered perturbation theory. We write the solution of 
$H_q$ as 
\begin{equation}
\psi(x) =\sum _\alpha u_\alpha(x)a_\alpha +
        \sum _\beta v_\beta (x)b^\dagger _\beta, \label{Hq}
\end{equation}
where $u_\alpha(x)=e^{-iE_\alpha t}u_\alpha (\vec x)\tau_\alpha$,
$v_\beta (x)=e^{iE_\beta t}v_\beta (\vec x)\tau_\beta$; $\tau$ is the
flavor wavefunction and the spatial wavefunction is:
\begin{equation}
u_\alpha (\vec x)=
\left(
\begin{array}{c}
g_{njl}(r) \\
-i \vec \sigma \cdot \hat{\vec r} f_{njl} (r)
\end{array} \right)Y_{jl}^m (\hat{\vec r}), \label{u}
\end{equation}
where $g$ and $f$ are real functions, $n$ is the radial quantum number,
and $Y_{jl}^m (\hat{\vec r})$ are the vector spherical harmonics. For
computational convenience, we will take the 
same form for $v_\beta(\vec x)$.

In correspondence to Eq. (\ref{Hq}), the quark propagator is
\begin{eqnarray}
D(x_1,x_2)&\equiv& \langle 0|T\{\psi(x_1),\bar\psi (x_2)\}|0\rangle \nonumber\\
        &=& \theta(t_1-t_2)\sum _\alpha u_\alpha(x_1) \bar u_\alpha(x_2)-
        \nonumber \\
        &&  \theta(t_2-t_1)\sum _\beta v_\beta(x_1) \bar v_\beta(x_2).
\end{eqnarray}

Applying the propagators to Fig. 1, we get the contribution for a single quark
line (with the initial and final states denoted as $u_i$ and $u_f$, 
respectively):
\begin{eqnarray}
\mu_s&=&\frac{1}{F_\pi^2}
        \int d^4x_1 d^4x_2 \bar u_f(x_2) S(r_2)\gamma ^5 \lambda^i
        \times \nonumber \\
        && \left[\theta(t_2-t)\theta(t-t_1)\sum _{\alpha\alpha'}u_\alpha(x_2)
        \Gamma_{\alpha\alpha'}
                \bar u_{\alpha'}(x_1)+ \right.\nonumber \\
        &&~ \theta(t_1-t)\theta(t-t_2)\sum_{\beta\beta'}v_\beta(x_2)
        \Gamma_{\beta\beta'}
                \bar v_{\beta'}(x_1) - \nonumber \\
        &&~\theta(t_2-t)\theta(t_1-t)\sum_{\alpha\beta'} u_\alpha(x_2)
        \Gamma_{\alpha\beta'}
                \bar v_{\beta'}(x_1) -\nonumber \\
        &&\left.
        \theta(t-t_2)\theta(t-t_1)\sum _{\beta \alpha'}v_\beta (x_2)
        \Gamma_{\beta\alpha'}
                \bar u_{\alpha'}(x_1)\right]\times  \nonumber \\
        && S(r_1)\gamma^5 \lambda^i u_i(x_1) \frac{i}{(2\pi)^4} \int d^4 k
                \frac{\delta _{ij} e^{-ik\cdot(x_1-x_2)}}
                {k^2-m_{\phi_i}^2+i\epsilon}
                \label{expr},
\end{eqnarray}
where
$\Gamma_{\alpha\alpha'}\equiv
        \int d^3x u_\alpha^\dagger (\vec x \times \vec \alpha)_3 u_{\alpha'}$,
and similarly for
$\Gamma_{\beta\beta'}$ etc.
The four time-ordered terms in Eq.(\ref{expr}) correspond to the
time-ordered diagrams of Fig. 2.

\begin{center}
\begin{minipage}{8.5cm}
\begin{figure}
\begin{center}
\psfig{figure=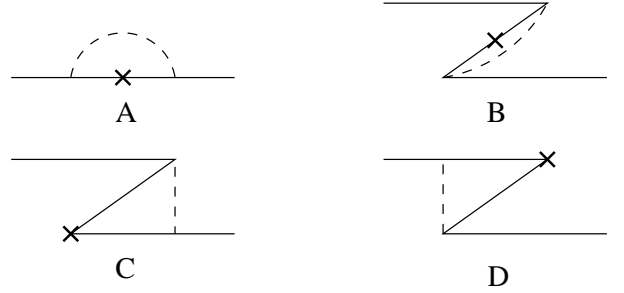,width=8cm}

\bigskip

\begin{tighten}
\caption{Time-ordered diagrams of Fig. 1; A is the positive-energy state 
contribution; B is the negative-energy state contribution; C and D are the 
quark-antiquark pair creation and annihilation ``$Z$'' diagrams.}
\end{tighten}
\end{center}
\end{figure}
\end{minipage}
\end{center}

We omit here the details for the evaluation of $\mu_s$ of Eq. (\ref{expr}). 
The integrals of Eq. (\ref{expr}) can be reduced analytically to
radial integrations at the vertex points ($r_1$ and $r_2$) and of the loop 
momentum $|\vec k|$. 
The the remaining integrations are carried out
numerically. To obtain $\mu_s$ for the whole nucleon, one still has to 
multiply a 
spin-isospin factor which can be straightforwardly calculated to be 2.

When calculating $\mu_s$ we allow strong variations of the model parameters 
entering in the Lagrangian fo Eq. (\ref{Lagrangian}). 
Since $F_\pi=93$ MeV and $m_K=495$ MeV are fixed by experiment,
our model contains four free parameters:
the two quark masses $m_{u,d}$, $m_s$ and the two strength constants
of the scalar and
vector potential denoted by $c$ and $\alpha$. The parameter $\alpha$
is fixed by the long-wavelength, transverse fluctuations of the QCD
based static-source flux-tube picture \cite{flux1,flux2}. 
It was obtained to
be $0.26$ in \cite{alpha1} and $0.30$ in \cite{alpha2}, while a much larger
value of about $0.52$ was used by the Cornell group \cite{Cornell}.
Recent lattice calculation \cite{Bali} got a value around $0.32$ in the
quenched approximation, and suggested that relaxing the quenched
approximation may lead to $\alpha \sim 0.40$.
Effective quark masses and confinement strength are rather uncertain 
quantities. 
We therefore choose in our calculation five different sets of parameters 
(see Table I),
including both current and constituent quark masses.

\begin{table}
\begin{center}
\begin{tighten}
\caption{Model parameters and the contribution to $\mu_s$ from the intermediate
quark ground state only.} 
\end{tighten}
\begin{scriptsize}
\begin{tabular}{cccccc}
para. &$m_{u,d}$ &$m_s$  & $\alpha$ & $c$ &$\mu_s$ [$\mu_N$] \\
set   &[MeV]&[MeV]&    &[GeV$^2$] &ground state\\ \hline
 1    &10 &150&0.26&0.11   &$-0.0115$\\
 2    &10 &150&0.26&0.16   &$-0.0176$\\
 3    &300&500&0.26&0.11   &$-0.0176$\\
 4    &10 &150&0.30&0.16   &$-0.0180$\\
 5    &10 &150&0.50&0.18   &$-0.0226$
\end{tabular}
\end{scriptsize}
\end{center}
\end{table}

Fig. 3 gives the numerical results of $\mu_s$ 
in units of $\mu_N$.
The intermediate quark states 
are consistently summed 
up to a given energy. In the last column of Table
I we list the contribution of the intermediate ground state. As evident from 
Fig. 3, 
for {\em all} choices of parameter sets, $\mu_s$ turns out to be 
{\em positive}, as 
long as enough excited quark states are taken into account. 
However, as already indicated in 
Table I the 
intermediate quark ground state always give a negative contribution. This 
explains why in many calculations at the baryon level, where the quarks are 
restricted to the 
ground state and the intermediate baryon is truncated to be the
ground state octet or decuplet baryons, a negative value of $\mu_s$ is usually 
obtained. 

\begin{center}
\begin{minipage}{8.5cm}
\begin{figure}
\begin{center}
\psfig{figure=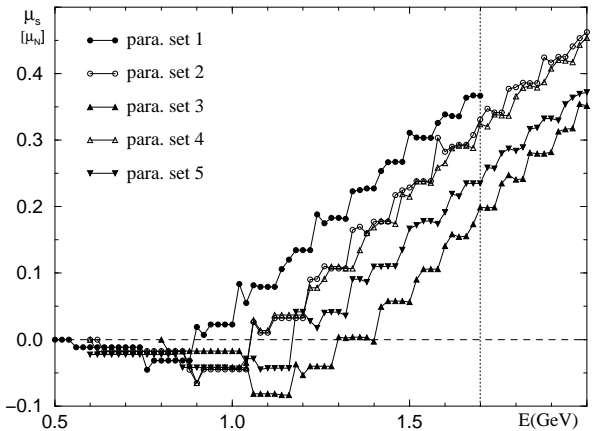,width=8cm}
\bigskip
\begin{tighten}
\caption{Plot of $\mu_s$ as a function of the maximal energy up to which the
intermediate states are summed.}
\end{tighten}
\end{center}
\end{figure}
\end{minipage}
\end{center}

We see in Fig. 3 that the summation over the quark intermediate 
states is divergent.
This is because in the chiral Lagrangian of Eq. (\ref{Lagrangian}), the 
electromagnetic current of the strange quark is not separately conserved. 
To obtain a meaningful finite result, we must renormalize the composite,
non-conserved magnetic moment operator of the strange quark. 
Analogous to the lattice renormalization, we cut the quark
intermediate states at a certain energy, which should roughly correspond to the 
inverse of the lattice spacing 
($a^{-1}\sim 1.7$GeV)  
in the lattice QCD calculation of nucleon 
properties \cite{a}. The cutoff point is indicated in Fig. 3. 

Fig. 3 shows that a larger quark mass or a 
stronger confinement (which is effectively a static mass) 
always reduce the magnetic moment, as expected. However, the 
variations of the vector potential do not affect $\mu _s$ too much.   

To analyze how the positive value for $\mu_s$ arises, 
in Fig. 5 we give the separate contributions to 
$\mu_s$ from
the time-ordered diagrams of Fig. 2 for the second set of parameters.
Correspondingly in Fig. 6 we indicate the separate contributions of the
time-ordered diagrams to strange quark polarization $\Delta s$ of the nucleon
\cite{Chen}. 

\begin{center}
\begin{minipage}{8.5cm}
\begin{figure}
\begin{center}
\psfig{figure=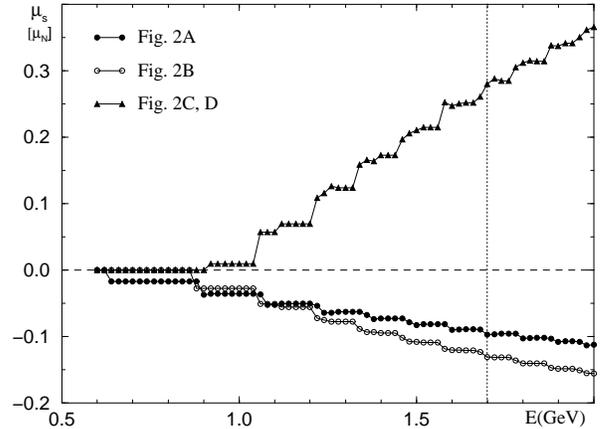,width=8cm}
\bigskip
\begin{tighten}
\caption{Contributions to $\mu_s$ from the time-ordered diagrams of Fig. 2;
the positive-energy and negative-energy states both generate a negative 
contribution, while the two ``Z'' diagrams yield a 
positive contribution. The results are given as a function of the energy 
cutoff on the intermediate quark states.}
\end{tighten}
\end{center}
\end{figure}
\end{minipage}
\end{center}

\begin{center}
\begin{minipage}{8.5cm}
\begin{figure}
\begin{center}
\psfig{figure=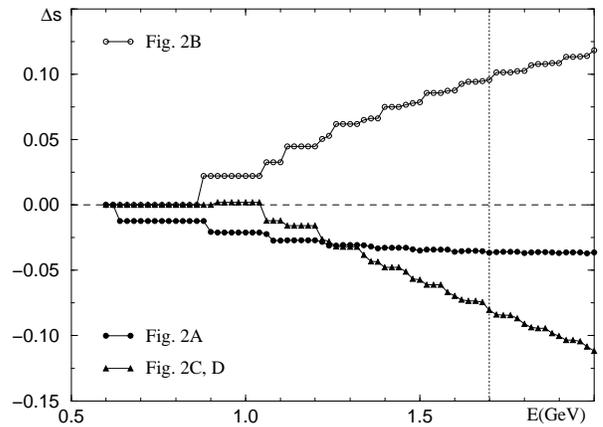,width=8cm}
\bigskip
\begin{tighten}
\caption{Time-ordered diagrams' contributions to $\Delta s$; 
the positive-energy,
negative-energy states, and the ``Z'' diagrams give a negative, positive, and 
negative contribution, respectively. The results are given as a function of 
the energy cutoff on the intermediate quark states.} 
\end{tighten}
\end{center}
\end{figure}
\end{minipage}
\end{center}

The results of 
Figs. 5 and 6 indicate why $\mu_s$ and $\Delta s$ 
have opposite sign: the intermediate quark states give a
contribution of the same sign (both negative) to $\mu_s$ and $\Delta s$, 
which is as expected; the antiquark states contribute a positive
amount to $\Delta s$, but a negative amount to $\mu_s$. This is also
reasonable since the antiquark has an opposite charge to the quark. 
However, one
would not expect the usual relation between spin and magnetic moment
for the contributions from the 
``Z'' diagrams in which a 
quark-antiquark pair is created or annihilated;
an evident obstacle is that
we do not know what sign of the charge we should attribute 
to these diagrams. Figs. 5 and
6 show that the ``Z'' diagrams give a negative contribution to the
polarization while they generate 
a positive contribution to magnetic moment. They are also 
the dominating contributions (note that there are two ``Z'' diagrams);
so eventually we get for the whole nucleon a negative strangeness polarization
but a positive strangeness magnetic moment. This agrees well with 
the experimental results. 

In summary, we found by a standard perturbative calculation that the SU(3) 
chiral 
potential model predicts a {\em positive} nucleon   
strangeness magnetic moment for a wide range of model parameters. Here
the contributions from the intermediate excited states and the ``Z'' diagrams
are most important. If one restricts the intermediate state to the quark ground
state an opposite result is obtained.  
Further investigation of the time-ordered
diagrams reveals that the positive-energy and negative-energy states 
(Figs. 2A and 2B) 
contribute to polarization and magnetic moment in the usual way that 
respects the relation between spin and magnetic moment;
however this is not true for the ``Z'' diagrams, whose 
contribution is found to be dominant and therefore determines the 
overall sign of $\mu_s$ 
and $\Delta s$ for the nucleon.  
Since in our calculations both for $\mu_s$ and $\Delta s$ we have 
assumed only one Lagrangian, the success of our calculations can be
regarded as a strong support of our approach and the chiral Lagrangian. 

This work is supported by the CNSF (19675018), CSED, CSSTC, 
the DFG (FA67/25-1), and the DAAD.

\end{multicols}
\end{document}